\documentclass[final,1p,times]{elsarticle}

\usepackage{amssymb}
\usepackage[raggedrightboxes]{ragged2e}

\usepackage [english]{babel}
\usepackage [autostyle, english = american]{csquotes}
\MakeOuterQuote{"}

\usepackage{color}
\newcommand{\rev}[1]{#1}

\journal{Space Policy}

\begin{document}

\begin{frontmatter}

\title{Can a Referendum Solve Problems of Shared Sovereignty on Mars?}

\author[inst1]{Roxanne Ruixian Zhu}

\author[inst1]{Jacob Haqq-Misra\corref{cor1}}
\ead{jacob@bmsis.org}
\cortext[cor1]{Corresponding author}


\affiliation[inst1]{organization={Blue Marble Space Institute of Science},Department and Organization
            addressline={600 1st Avenue, 1st Floor}, 
            city={Seattle},
            state={WA},
            postcode={98104}, 
            country={USA}}

\begin{abstract}
Space exploration technology continues to expand humanity's reach beyond Earth, and even more ambitious efforts are striving to establish long-duration human settlements on Mars. The dependence of martian settlers on life-support infrastructure and on resupply missions from the host nation could create conditions for tyranny or lead to other extreme and uncontrollable situations, but such risks could be reduced by thinking about the possibilities for effective decision making on Mars before any settlement efforts actually occur. This paper examines the extent to which referendums could be used on Mars as a means of political decision-making and sovereignty adjudication. Our approach draws on three terrestrial case studies---the Great Idaho Movement in the United States, the Catalan Independence Movement in Spain, and the Quebec Independence Movement in Canada---as potential analogs for Mars governance. We recommend advance determination of the conditions under which a martian referendum would be recognized as a best practice for any agency seeking to establish a long-duration settlement on Mars. We suggest that referendums can reduce the likelihood of multiple authoritative political entities existing on Mars, which could provide a more procedural approach toward resolving governance issues between Earth and Mars. However, if Mars settlement is successful in establishing settlements on the scale of cities or larger, then other uniquely martian tools may evolve as a supplement or replacement to referendums.
\end{abstract}



\end{frontmatter}

\newpage


\section{Introduction}
\label{sec:introduction}

Ongoing efforts to explore and commercialize space by state and private agencies are further enabling the capacity for human spaceflight, which include technology development that could enable the long-term or permanent human settlement of space. Several nations have enacted national legislation that authorizes forms of space mining, including the United States, Luxemburg, Japan, and the United Arab Emirates, which are intended to support the emergence of private industries that seek to harvest resources such as water, rare earth elements, and precious metals from nearby asteroids. Longer-term ambitions for establishing human outposts on Mars are being pursued by NASA, SpaceX, the China National Space Agency, and the United Arab Emirates, although the timeline for sending the first humans to Mars remains uncertain. Any actual attempts at sending humans to Mars for short-duration visits will likely face unanticipated challenges, as will any subsequent efforts to establish continuously-inhabited outposts. Current efforts may even be disrupted by political or economic conditions and delayed until some unforeseen time in the future. And yet, the fact that current technology can conceivably enable the human settlement of space \cite[see, e.g.,][]{zubrin2011case,musk2017making} heightens the importance of thinking about effective strategies for space settlement long before the first human sets foot on an asteroid or Mars. As the technological capacity for space settlement continues to develop, human systems of governance must likewise consider the implications of expanding the domain of terrestrial political entities beyond Earth to encompass space resources and even another planet.

Long-term visions of space settlement on remote asteroid outposts and especially on Mars can raise novel thought experiments in ethics and governance that may be unlikely to arise or overlooked in a typical terrestrial setting. For example, breathable air will be a limited resource for any human settlers on Mars—one that must be technologically manufactured with an uninterrupted supply for each person—so the ability to produce breathable air in sufficient quantity, at affordable costs, and with sufficient redundancy to mitigate risk will place physical limitations on the population size that can be supported \cite{stevens2015price}. Attempting to live in the space environment raises numerous challenges that derive from the increased dependence of human survival on technically managed infrastructure, and the corresponding risks that arise for intentional or unintentional mismanagement that could undermine basic freedoms \cite[e.g.,][]{cockell2014meaning,cockell2015human,cockell2016dissent,cockell2022interplanetary}. Processes for making decisions for isolated groups of humans living in space also serve as novel thought experiments in governance, which can range from ideas for ensuring effective collective decision making in the harsh environment of space to models that could enable partial or full autonomy for a settlement in space \cite[e.g.,][]{gilley2020space,froehlich2021assessing,haqq2022sovereign,cockell2022institutions,nesvold2023off,schwartz2023reclaiming}. The value of considering space as a domain for such thought experiments is twofold. First, space settlement is actually being attempted, even if it is a long-term goal, so thinking about governance strategies for space today can help to prepare for this eventual future. Second, the space domain is a venue for probing the boundaries of ideas that may otherwise be taken for granted, which can provide novel insight for existing ideas on Earth. Space settlement may be a long way off, or may never occur, but thinking about space settlement can be useful for understanding our own future on Earth. 

Collective decision making among communities living together in space remains an important consideration for such thought experiments. The harsh reality of the space environment will sometimes require rapid decisions that cannot always be subjected to a group vote; but at the same time, the possibility that a malevolent actor or group of actors could leverage control of technical infrastructure for self-serving political purposes is only one example that highlights the need for political decision-making strategies that can avoid such tendencies toward tyranny. Other scenarios may involve settlers seeking to voice concern in unity, perhaps targeted toward their host nation or another nation operating on Mars, with the goal of influencing geopolitics or raising awareness of an important issue. Even more grandiose ideas could involve a political movement for liberation or secession, in which a settlement on Mars achieves technical and political self-sufficiency to the extent that it declares itself independent from political entities on Earth. In any of these cases, historical analogs from Earth’s history provide a basis for thinking about possible situations of human governance in space. The use of history as an analogy does not necessarily imply that the future will follow historical patterns, and caution should be taken when attempting to draw parallels from analogy (see, e.g., the discussion by \citet{dick2018astrobiology}). But historical analogy can also provide insight into solutions that are known, as well as those that have been attempted and failed on Earth, which can be a helpful place to begin thinking about human futures in space.

This article examines the concept of a voter referendum as a mechanism for political decision making and the determination of sovereignty on Mars. The specific focus is on the extent to which referendums could enable a community of settlers on Mars to exert autonomy or declare independence from its host nation, based on a comparative analysis of three terrestrial case studies: the Quebec Independence Movement, the Catalan Independence Movement, and the Greater Idaho Movement. \rev{The Quebec Independence Movement is an example of a referendum conducted within a legal framework that was successful to an extent in gaining autonomy for Quebec, even though it did not lead to full independence from Canada. The Catalan Independence movement is an example of a referendum that was counter to the existing legal paradigm and that ultimately vailed. The Greater Idaho movement seeks to shift political boundaries between states in the United States, which is an ongoing effort that is embedded within existing legal frameworks.} These examples of recent voter referendums toward altering existing sovereign relationships can provide insight into \rev{aspects of} thinking about governance solutions that may be possible in space.

\section{Referendum Case Studies}

The history of referendums can be traced back to the Athens Assembly in ancient Greece (\textit{ecclesia}), which was depicted in the plays of Aristophanes, an ancient Greek comic playwright. Citizens of the Athens Assembly enjoyed the right to speak, discuss, and resolve city-state affairs, although eligibility was restricted to male Athenian citizens over the age of 20. The practice of ostracism, in which a popular vote could sentence even powerful individuals to ten years of exile, was a notable invention and representative example of this democratic body, marking the earliest historical appearance of the referendum \cite{develin1978provocatio,hammer2005plebiscitary}. After the fall of the Roman Empire, the referendum system gradually declined. However, the universally acknowledged first referendum in history occurred during the French Revolution, known as the Avignon Referendum, where over 100,000 out of 153,000 voters supported Avignon's separation from papal control and its incorporation into France \cite{morel2021}. Following World War I, President Woodrow Wilson of the United States, drawing inspiration from the "independence referendum" concept introduced during the Avignon Referendum, proposed the principle of "national self-determination," supporting groups of people to form their own states and governments during reconstruction. Throughout the 20th century, referendums emerged as one of the most vibrant forms of democracy alongside representative democracy. Compared to indirect democracy, referendums can often be more reflective of the people's positions and attitudes. As real-world issues grow in complexity and diversity, and disagreements among policymakers intensify, leading to decreased efficiency in representative democracy, the maturity of the referendum system can fill some gaps and optimize power structures.

Referendums have been utilized on numerous occasions in adjudicating issues of sovereignty. In modern times, with the vigorous rise of national liberation movements worldwide, over one hundred former colonies and dependencies have successively gained independence. A nationwide referendum for independence represents the will of the local residents and has been employed as a legal tool in decolonization. The Liberian movement in 1846-1847 relied on clear constitutional and legal grounds to achieve independence after the referendum's approval reached the statutory proportion. In 1846, the US Congress passed the Act of Incorporation and Government of Liberia, ending direct rule over Liberia. On July 26, 1847, the Federal Republic of Liberia convened a constitutional convention, adopted its Declaration of Independence and Constitution, established its national flag and coat of arms, and officially established the independent Republic of Liberia. Similarly, the Republic of Maryland achieved independence in 1853 through its own referendum process, although the Republic of Maryland merged with Liberia in 1857. This represents one of the earliest successful cases of independence achieved through a referendum with clear legal basis in modern times. The referendum on the dissolution of the Sweden-Norway Union in 1905 was also successful after Norway's authorization and consent, leading to Sweden officially recognizing Norway's independence on October 26, 1905. The referendum on the independence of Outer Mongolia in 1945 relied on the Sino-Soviet Treaty, which stipulated that independence would be achieved after the referendum's approval reached the statutory proportion. These examples show that the theory of independence referendums has a long history and has been increasingly utilized in modern times as colonialism disintegrated, national sovereignty consciousness emerged, and international law developed. Rooted in Western history, referendums have become an important supplement and alternative to representative democracy in modern times, and can be recognized as legally binding international law acts when grounded in constitutional provisions.

The remainder of this section will discuss the Quebec Independence Movement, the Catalan Independence Movement, and the Greater Idaho Movement as recent examples of referendums that have been attempted to change the relationship between existing sovereign entities. \rev{These case studies were all selected because of their relevance as positive or negative examples of referendums that have some relevance to future applications in space. The Quebec Independence Movement has its origins in strong cultural divisions between those living in Quebec and other Canadian provinces, which led to a recognition by the Canadian government of a legal mechanism for voter independence referendums along with other concessions of autonomy to Quebec. Such a scenario could be relevant to a settlement on Mars that may have its own strong cultural unity and desire to preserve aspects of its autonomy, even while maintaining a sovereign relationship with a host nation on Earth. The Catalan Independence Movement was similarly motivated by the desire of those living in Catalan for greater autonomy from Spain, but this effort was contrary to the provisions of the Spanish Constitution that bound Catalan, so the independence movement ultimately failed. Similar scenarios could play out in a settlement on Mars that may seek to assert its own independence or autonomy that is contrary to preexisting agreements or treaties that were in place at the time the settlement effort began. The Greater Idaho Movement does not seek independence but instead is an attempt to redefine political boundaries between states in the United States, due to the greater cultural affinity between parts of eastern Oregon and the state of Idaho. Similar political dynamics could arise in a situation in which multiple nations have established settlements on Mars, but certain groups of settlers begin to express a desire to realign their sovereign ties to a different host nation.} None of these examples have been wholly successful, but the details of each case study can be instructive in thinking about how such referendums could be successful in the modern space age. 

\subsection{Quebec Independence Movement}

Quebec was originally a French colony in North America, where the French transplanted their history and culture and spent nearly two hundred years. The Seven Years' War between Britain and France broke out in 1754, and France suffered a crushing defeat in 1763, signing the Treaty of Paris, which ceded all its colonies in Canada to Britain. In the same year, British Canada was officially established. The British government pursued an "Anglicization" policy in Canada, aiming to establish British traditional political systems, laws, languages, and religious beliefs in Quebec. However, the French-Canadian residents on this land still identified themselves with their past, preserving their French cultural traditions. Quebec even continued to adopt French laws, and the British "assimilation policy" encountered strong resistance from the French-Canadian residents. Consequently, the Quebec independence movement emerged, experiencing a lengthy period of transition, disputes, and compromises, culminating in voter referendums in the 1980s and 1990s.
During British rule, various acts such as the Quebec Act, the Canadian Act, the Union Act, and the Bill of Rights were enacted to reconcile the conflicts between Quebec and the federal government. Although the Bill of Rights in 1840 ultimately united Upper and Lower Canada (divided by the Ottawa River in 1791, with Upper Canada adopting the British legal system and Lower Canada remaining unchanged) and established the Province of Canada, politically unifying English and French Canadians, the rift between the two nations remained difficult to erase. Canada's long allegiance to Britain led to undercurrents between the central and local governments.

After World War II, Quebec's economy underwent a transformation from a traditional agricultural society to a modern one. The state apparatus was reorganized, social services nationalized, and Catholic influence rapidly declined. However, along with industrial progress, social issues deepened. The fact that a significant portion of Quebec's economy was controlled by Anglophone residents sparked discontent among French-Canadians, who sought equal political and economic rights comparable to those of Anglophone Canadians. Additionally, the influence of neighboring Ontario on Quebec gradually intensified, eroding its advantageous river and transportation positions. It was against this backdrop that the "Quiet Revolution" erupted. In 1960, René Lévesque founded the Parti Québécois and became Premier of Quebec in 1976, initiating a comprehensive reform. During the Quiet Revolution, the Liberal provincial government worked to diminish the Church's influence on Quebec's education, established the Ministry of Culture, enacted the Charter of the French Language, establishing French as the sole official language in Quebec. Economically, it shifted away from a policy of strict liberalism, reclaiming economic power previously held by Anglophone and Anglo-American capitalists for the Quebec government and enhancing the status of French Canadians through various measures. In the political realm, the "Sovereignty-Association" framework was proposed, advocating economic ties with the federation while pursuing political sovereignty.

In 1980, Quebec held its first provincial referendum on sovereignty, voting on the "Sovereignty-Association" platform. Of the voters, 59.56\% opposed secession from the federal government, and the first independence referendum by the Quebecois ended in failure. In 1995, three Quebec independence-advocating parties (Parti Québécois, Rassemblement pour l'Indépendance Nationale, and Nouveau Parti Démocratique du Québec) initiated the second referendum on independence, with the unity faction emerging victorious again by a narrow margin of 1.16\%, marking the second referendum's failure and once again narrowly averting the country's fragmentation. Undoubtedly, following the two referendums, the federal government recognized the gravity of the situation. The day after the second referendum, Prime Minister Chrétien vowed to promote constitutional reform supported by other provinces and endeavor to persuade them to acknowledge Quebec as a "distinct society." Faced with the looming threat of secession, Canadians questioned the government, inquiring why Quebec could unilaterally hold referendums and set referendum themes without any domestic or international legal basis. Amid escalating societal debates, the Canadian federal government ultimately resorted to legislation, enacting the Clarity Act and Bill 99, which affirmed Quebec's right to national self-determination and the authority to determine its political system and legal status, yet stipulated that a negotiated referendum under the Clarity Act with a 50\%+1 vote outcome would be necessary for such decisions. On November 27, 2006, the Canadian House of Commons passed a resolution put forward by then-Prime Minister Stephen Harper with 226 votes in favor and 16 against, acknowledging that "the Québécois form a nation within a united Canada" \cite{canadagovt}. By devolving power, Harper's resolution, albeit unprecedented in making a national prime minister and parliament recognize a province as a nation, somewhat calmed the Quebec independence movement. Indeed, the Quebec independence movement can be considered partially successful, as Quebec continues to exist as a "distinct society," and Canada has successfully averted national fragmentation.

\subsection{Catalan Independence Movement}

Another illustrative case is the Catalan independence movement in Spain. Catalonia is one of the most economically developed regions in Spain. Although its population accounts for only 16\% of the total, it contributes more than 20\% of the country's tax revenue, fostering lingering dissatisfaction among Catalans towards the Spanish central government. Beyond economic issues, the Catalan independence movement is also rooted in historical factors, as Catalans seek not only economic parity but also political self-determination. Catalonia has maintained a highly autonomous government system for an extended period. In 1469, Catalonia officially joined the Spanish Habsburg dynasty alongside Aragon, yet its parliament was not abolished but rather retained as a constitutional local autonomy. From 1640 to 1650, Catalonia's divergent stance during the Spanish Succession War sparked a major separatist uprising, leading to the establishment of the Catalan Republic, which was reconquered by Spain in 1714 and retained within Spain under the name of Catalan Autonomous Government. From the post-World War I era to the present, Catalonia has continuously sought independence through negotiations and confrontations with the Spanish government.

However, unlike the Canadian federal government's approach, the Spanish government has refused to recognize Catalonia's self-determination. A referendum held in 2006  approved the "Statute of Autonomy of Catalonia" by Catalonian voters, but the legitimacy of these statues were challenged by the Spanish government from the outset, who deemed Catalonia’s attempt at national self-determination through referendum as illegal. During another referendum in Catalonia in 2017, local authorities declared victory with a 90\% support rate, prompting the Spanish central government to send police forces to suppress the vote, effectively declaring the referendum a failure. The Spanish central government argued that Catalonia's autonomy stems from the Spanish Constitution, which grants it its rights. It is evident that the "Statute of Autonomy of Catalonia," adopted through a local referendum, falls under the jurisdiction of the Spanish Constitution. The conditions for a referendum are provided in Section 92 of the Spanish Constitution, which stipulates that
\begin{quote}
(1) Political decisions of special importance may be submitted to all citizens in a consultative referendum.\\
(2) The referendum shall be called by the King on the President of the Government's proposal after previous authorization by the Congress.\\
(3) An organic act shall lay down the terms and procedures for the different kinds of referendum provided for in this Constitution.    
\end{quote}
Catalonia's unilateral adoption of an independence referendum without the consent of the nation and the central government violates the Constitution, which thereby justifies the central government's countermeasures. To a certain extent, this fundamentally thwarts Catalonia's aspirations for independence, compelling it to pursue other avenues for enhancing autonomy.

\subsection{Greater Idaho Movement}

The final case is the Greater Idaho Movement in the United States, whose future is still unknown. The Greater Idaho movement initially emerged in 2020 within two rural counties of Oregon and swiftly expanded to encompass other neighboring counties. The movement primarily focuses on the eastern counties of Oregon, bordering Idaho. These regions, being predominantly rural, have seen a surge of discontent among the local populace towards policies enacted by the Democratic Party's administration in Oregon, particularly the legalization of hard drugs and the subsequent escalation of social crime rates. Consequently, a majority of the residents have shifted their allegiance towards the Republican Party. However, with the Democratic Party maintaining a dominant position in Oregon, while the Republican Party enjoys more support in adjacent Idaho and other areas, a myriad of factors have intensified political divisions within Oregon. The rift between the sparsely populated, relatively conservative eastern rural areas and the liberal, densely populated urban centers such as Portland and the state capital of Salem has become increasingly pronounced.

The fundamental aim of the Greater Idaho movement is to redress this imbalance of political power distribution through the alteration of state boundaries, thereby better reflecting the interests of local residents. Its primary demand is the annexation of 17 counties from Oregon into Idaho, encompassing the entirety of 14 counties and portions of 3 others. This action would shift Idaho's border westward by approximately 200 miles, traversing the heart of Oregon. Supporters argue that such a move would not only preserve the traditional values of rural communities but also integrate them into a state that more effectively represents their interests---Idaho, where the Republican Party holds a majority.

Since its inception, the Greater Idaho movement has garnered widespread support from many residents within the eastern counties. The movement progressed rapidly in its first two years, garnering votes from a total of 11 counties out of 36 in Oregon by the end of 2022. The movement then grew more slowly. In the vote held on May 28, 2024, Crook County became the latest county to support the Greater Idaho Measure. Measure 7-86 passed in Crook County with 53\% of the vote, injecting new life into the Greater Idaho movement.

Commentators within the United States have noted that Oregon has a history of adopting novel policies before their full social impact is understood, and the Greater Idaho movement is perhaps another instance of Oregon flirting with unconventional solutions. While the movement has made some progress, it has yet to achieve its ultimate goal. A nonbinding measure was passed by the Idaho House in 2023, indicating receptivity to talks between Oregon and Idaho regarding the possible relocation of the state borders.  Nevertheless, the future direction and outcome of this movement remains uncertain.

\section{Applications to Mars}

Historical examples of referendums within modern nation-states can provide insight into the application of similar political tools for human settlements in space. Cases of successful referendums can suggest factors that may be relevant when considering the implementation of future referendums, on Mars or elsewhere, and cases of failed referendums can likewise indicate broader tendencies that may also apply in the space environment. Historical precedent does not serve as a predictor of future human behavior, especially in space, but these historical cases can serve as a basis for thinking about the factors that would be relevant for any attempt at implementing referendums for populations living in space.

The three cases examined in this study represent distinct approaches and outcomes of the use of referendums to seek political or economic independence, summarized in Table \ref{tab:referendums}. The Quebec independence movement was inherently driven by a strong sentiment for autonomy from the outset. Despite failing to achieve sovereign independence in two referendums, the Quebec independence movement ultimately secured recognition from the Canadian federal government as a "nation within a nation," existing as a unique society and enjoying much more political autonomy compared to the other two cases. Conversely, the Catalan movement was deemed "unconstitutional" by the Spanish government from the beginning, as its independence aspirations carried strong risks of national fragmentation and faced opposition from much of the Western world. The national government, holding the legal high ground, took firm measures to ensure its own territorial integrity, eliminating the possibility of Catalan autonomy. Unlike these two movements, the Greater Idaho Movement's referendum was politically legitimate and not a traditional secessionist act. Rather, this movement focused on the transfer of state borders due to the political and cultural affinity of counties in eastern Oregon with Idaho, and thus holds the potential for success as an ongoing issue. Furthermore, the first two cases were rooted in profound historical and cultural conflicts and marked by sharper economic disparities. Catalans felt they contributed disproportionately large taxes without commensurate returns, while French Canadians believed they were oppressed by the Anglophone majority. In contrast, the Greater Idaho Movement's narrative is based on political differences between the two major US parties, rather than economic issues in the region.

\begin{table}[h!]
\centering
\scriptsize
    \begin{tabular}{p{0.25\linewidth}|p{0.35\linewidth}|p{0.1\linewidth}|p{0.15\linewidth}} 
    \hline
    \textbf{Case Study} & \textbf{Main Demands} & \textbf{Legality} & \textbf{Result} \\
    \hline\hline
    Quebec Independence Movement & Seek political independence but maintain economic ties & Legal & Successful (to some degree) \\
    Catalan Independence Movement & Seek both political and economic independence & Illegal & Failed \\
    Greater Idaho Movement & Seek shift of state political borders, not full sovereign independence & Legal & Ongoing \\  
 \hline
\end{tabular}
\label{tab:referendums}
\caption{Overview of the three referendum case studies.}
\end{table}

Drawing from these three cases, we can delve into the possibility of employing referendums for sovereignty decisions on Mars.  On Mars, it may be possible to avoid the deep-seated historical conflicts that plagued the Quebec and Catalan independence movements. Martian settlers will essentially be developing a new culture that draws from various ethnic backgrounds, each carrying distinct cultural heritages—although the ethnic composition of any settlement may vary based on the nation that initiates the mission. The first martian settlements will likely have small initial populations, so it will be challenging to form significant cultural enclaves \cite[e.g.,][]{salotti2020minimum,arguello2023exploration}. As the capacity for increased population expands, then the use of referendums may become more feasible as a means of addressing specific ongoing or arising issues that affect everyone. In this context, any referendums in the early stages of Mars settlement are most likely to align with the Idaho Freedom movement, with the goal of addressing rising political or cultural tensions within the framework of existing sovereign rule, thereby avoiding significant separatist tendencies and the resultant social unrest. 

If short-term settlement efforts are successful, then later stages could result in much larger population centers on Mars \cite[e.g.,][]{cannon2019feeding} that could result in scenarios where referendums are used for stronger political purposes of achieving political autonomy, as in the cases of Quebec and Catalonia. However, the ability for a settlement on Mars to achieve political and economic self-sufficiency is a daunting task, given the limited resources available on Mars and the initial dependence on supplies from Earth. The idea that a settlement on Mars might declare itself to be independent by referendum is perhaps conceivable in a distant future, but reaching this point of self-sufficiency would only have resulted from the long-term investment of infrastructure from national or private space agencies. If such resources were provided by a visionary benefactor for deep altruistic reasons, then long-term ambitions for an independent Mars may be feasible \cite[e.g.,][]{haqq2019can}; however, the lack of such intergenerational foresight suggests that short-term interests may dominate the settlement of Mars. Given the investment that spacefaring states will have made to enable a long-term human presence on Mars, such states may be reluctant to cede sovereignty over any of its infrastructure in space and, like the case of Catalonia, may refuse to recognize any referendum that calls for martian independence. 

\rev{It is noteworthy that the initial population of a martian settlement may range from several thousand to a hundred thousand individuals---far below the million (or greater) person thresholds typical of terrestrial sovereignty entities. However, Mars settlements will face extreme resource dependence (e.g., oxygen recycling in closed ecological systems, mineral reserves) that will amplify resource allocation conflicts among communities. For instance, the "economic disparities" observed in Quebec's case (Table \ref{tab:referendums}) might manifest as "oxygen quota disputes" or "energy production responsibility allocation" issues on Mars. While terrestrial sovereignty movements primarily involve the redistribution of political rights, emerging sovereignty movements in martian settlements could confront the intertwining of survival rights and autonomy at an earlier stage. Continued population growth in martian settlements could push life-support systems toward critical load thresholds, which could cause fiscal imbalances akin to Catalonia's "tax contribution-return disparity" that could manifest as "energy output-consumption quota controversies" on Mars that may even involve the compulsion of communities to accept technocrat-led central coordination mechanisms. Other examples of potential conflict scenarios on Mars that draw upon the three case studies are listed in Table \ref{tab:comparison}. Such considerations suggest the need for governance policies on Mars to establish a "Minimum Viable Sovereign Unit" model tailored to the specific martian contexts. This approach would consider the population density thresholds at which referendum mechanisms become operationally feasible; this would be similar to terrestrial precedent, but thresholds on Mars would be notably lower than those on Earth. Other considerations could include innovation in early warning metrics to prevent future conflict: for example, traditional economic disparity indicators (e.g., GDP ratios) could be supplemented by a Life Support System Contribution index, which would quantify marginal contributions of individuals or communities to vital systems such as oxygen production and water recycling. Metrics like these could serve as a parameter to identify resource or economic disparities among communities on Mars prior to the emergence of any conflict.}

\begin{table}[h!]
\centering
\scriptsize
    \begin{tabular}{p{0.30\linewidth}|p{0.35\linewidth}|p{0.25\linewidth}} 
    \hline
    \textbf{Earth Case Features} & \textbf{Adaptation to Martian Context} & \textbf{Potential Conflict Scenarios} \\
    \hline\hline
    Quebec: Cultural autonomy demands & Transformation of linguistic-cultural identity into technological-ecological identity & Disputes over genetic repository governance, AI ethical standard disagreements \\
    Catalonia: Legal confrontation & Supersession of terrestrial constitutional frameworks by interplanetary settlement accords & Arbitration of orbital resource extraction rights, extraterrestrial pollutant emission standards \\
    Greater Idaho: Legal boundary adjustment & Geographic boundaries replaced by orbital parameters and landing zone access rights & Negotiations over radiation shelter access permissions, conflicts during subterranean city expansion\\  
 \hline
\end{tabular}
\label{tab:comparison}
\caption{Examples of adaptations and conflict scenarios on Mars, based on the three referendum case studies.}
\end{table}

The best possibility for any approach to shared sovereignty on Mars would be to define the conditions that would be mutually recognized for independence or partial autonomy, ideally even prior to the establishment of any long-term human habitation in space. Regarding initial policy formulation, the Canadian federal government's approach is arguably the most insightful to emulate. Its referendum law, to some extent, precluded Quebec's path to independence while ensuring the legitimacy of the process. Eventually, the federal government also devolved some powers, leading Quebec to abandon its pursuit of complete independence and agree to sign the federal constitution. For Mars, this approach could enable a host nation to reduce the risk of secession by providing a pathway for semi-autonomous governance of a settlement that remains under the jurisdiction of the host nation. The implementation of a referendum in this context reflects the sentiments of the martian population for self-rule but without causing irreparable or uncontrollable impacts on the state. 

As a worst case scenario for Mars, the failure to preemptively consider the possibility that martian settlers may eventually desire degrees of political or economic independence may lead to situations in which a host nation must forcibly respond to attempts at secession from self-sustaining Mars settlements. In the case of Catalonia, the Spanish government suppressed the Catalan independence movement through extremely forceful means, threatening that Catalonia would be ostracized by Europe if it seceded. If a similar situation were to occur with a settlement on Mars, then intervening with direct military action would be much more challenging given the travel time between Earth and Mars: travel is generally only feasible every twenty-six months when the two planets are closest in orbit. Responding to undesired declarations of secession by a Mars settlement may instead take the form of sanctions, disabling of satellites, cyberwarfare, and other remote strategies for engaging in interplanetary conflict. Such a scenario may be extremely unlikely, due to the fragility of sustaining a settlement on Mars, and the extent to which a settlement on Mars could achieve full economic self-sufficiency remains uncertain.

\subsection{Legal Considerations}

Under international law, national territorial sovereignty in space is constrained by the Outer Space Treaty (formally the Treaty on Principles Governing the Activities of States in the Exploration and Use of Outer Space, including the Moon and Other Celestial Bodies) of 1967, which has been ratified by all spacefaring nations. Article II of the Outer Space Treaty states that Mars, like other celestial bodies, “is not subject to national appropriation by claim of sovereignty” whether by occupation or by any other means. This means that no territory on Mars or elsewhere in space can be incorporated into the sovereign geographical domain of a nation on Earth. However, Article VIII of the Outer Space Treaty specifies that states “retain jurisdiction and control” over objects and personnel launched in space or placed on a celestial body. This means that space stations and other permanent installations—and the people living in them—remain under the legal jurisdiction of their host nation while in space, even if the planetary surface that they occupy cannot be claimed as sovereign territory. Likewise, Article VI makes states responsible for the activities of non-governmental entities in space, with states required to provide “authorization and continuing supervision” for all activities by governmental and nongovernmental agencies. This means that even private space agencies are bound by their host nation’s commitment to the Outer Space Treaty, with any private property or employees launched into space still subject to the domestic laws of their host nation. Even with the Article II prohibition on explicit sovereign claims to territory in space, ongoing developments in space exploration and settlement will continue to expand the possibility of situations of legal conflicts that arise from sovereign jurisdiction of infrastructure in space or negligence of a nation to effectively oversee the activities of its private entities. 

Within the confines of the Outer Space Treaty, numerous possibilities remain for how an emerging settlement on Mars may develop systems of localized government. Although some idealistic possibilities have been suggested for how such systems could be constructed \cite[e.g.,][]{haqq2022sovereign,schwartz2023reclaiming}, the most likely outcome is that people arriving on Mars will inevitably carry their preexisting notions, making it challenging to develop novel governance structures or legal principles from the outset. Therefore, the jurisdiction over the physical infrastructure of any settlement on Mars will remain with the host nation, and individual host nations may differ in the extent to which they will recognize the legitimacy of any emerging governance structures on a Mars settlement. This makes it highly probable that any emerging governmental entities on Mars will be developed in consultation with the host state rather than in defiance of the host state’s sovereignty. 

Sovereignty issues could also arise in situations of shared infrastructure on Mars, where multiple states or agencies cooperate in the management of a large settlement or space station. Such situations already occur on the International Space Station, where each component that has been built and launched separately retains its jurisdiction under its host state; this leads to situations where the legal regime between different modules on the International Space Station actually changes based on where an astronaut, piece of equipment, or packet of data is located at a given time \cite{sadeh2004technical}. The International Space Station is maintained through a series of agreements that allow for such logistical challenges to be solved, and similar approaches could be used if shared infrastructure were built on Mars by multiple nations. Referendum issues could arise in shared infrastructure on Mars due to competing political or cultural desires, similar to the Greater Idaho movement case. Rather than seeking full autonomy, such cases could involve communities living in sections of shared infrastructure seeking to reassign their sovereign jurisdiction from one partnering nation to another. The Greater Idaho movement remains ongoing, but it seems likely that Oregon will object to any serious legal steps that are taken toward reassignment of any of its counties to Idaho, as the state of Oregon will seek to preserve jurisdiction over its territory. Likewise, any attempt at reassignment of any sections of settlement infrastructure on Mars from one nation to another would likely be opposed by the nation that would stand to lose jurisdiction over its property and citizens. In cases where such reassignment of jurisdiction makes sense, then a referendum might the first step toward signifying the desire for such a change, but the most pragmatic approach toward implementation would be by diplomatic agreements or treaties between the two nations.

\rev{The constitutional crisis triggered by the Catalan independence referendum and the jurisdictional conflicts within the space legal system fundamentally reveal the deep-seated contradictions between sovereign attribution and procedural justice in modern legal frameworks. The core logic behind the Spanish Constitutional Court's ruling that the Catalan referendum was unconstitutional lies in the fundamental conflict between its assertion of "Catalan popular sovereignty" and Article 2 of the Spanish Constitution, which establishes "the sovereignty of the Spanish people as a whole." This type of sovereign dispute would manifest in even more complex forms in the space domain. For example, Article II of the Outer Space Treaty explicitly prohibits states from claiming sovereignty over celestial bodies, but actions by private entities like SpaceX to appropriate resources through forms of "jurisdictional control" could threaten to undermine this non-appropriation principle. The similarity of these legal dilemmas resides in the inadequate adaptability of existing legal frameworks to accommodate claims from emerging actors.}

\rev{The Spanish government's invocation of Article 155 to suspend Catalan autonomy exposed the coordination failures of multi-tiered governance systems. This type of central-regional power struggle could escalate into sharper conflicts between national interests versus global public goods in space governance. For example, the international resource development regime outlined in Article 11 of the Moon Agreement has remained a dead letter due to major spacefaring nations' refusal to ratify the agreement. As a result, the current space mining legal situation is mired in a predicament akin to the Catalan crisis: existing international treaties fail to constrain new actors, while rule-making is hindered by great power politics. For instance, the U.S.-led Artemis Accords attempt to establish a "first-come, first-served" club model for lunar mining, which potentially could conflict with the Outer Space Treaty's prohibition on national appropriation. This legal vacuum mirrors the institutional deadlock between Catalonia's constitutional amendment rigidity (requiring a two-thirds supermajority in both chambers) and its separatist aspirations.}

\rev{Resolution pathways for both crises hinge on the dynamic adaptive capacity of legal systems. In its 2017 judgment, the Spanish Constitutional Court creatively introduced the "integrity of constitutional order" principle, emphasizing that any self-determination claims must pursue constitutional amendments through Article 168 procedures---thereby preserving possibilities for institutional evolution. For space governance, a similar approach could be to construct a "transitional legal framework,", which could include freezing sovereign disputes through interim agreements, establishing an international resource regulation authority, and prohibiting substantive possession until technological maturity. However, the situation in Catalan reached an impasse, so institutional innovations would be required to transcend such situations in space. One approach could be restructuring the United Nations Committee on the Peaceful Uses of Outer Space (COPUOS), endowing it with binding norm-making authority and dispute resolution mechanisms, akin to how Article 161 of the Spanish Constitution grants final adjudication power to the Constitutional Court. Ultimately, existing institutional arrangements or the development of new systems will be needed to maintain legal stability while opening institutional pathways for legitimate claims from emerging actors in space.}

\subsection{Semi-independent Settlements}

\rev{Any actual attempts at long-duration space settlement will lead to the emergence of international relations Between Earth and Mars and raise new questions about interplanetary cooperation or conflict. This subsection discusses the emergence of such dynamics between Earth and Mars based on the balance of power theory and hegemonic stability theory from international relations. The theory of the balance of power suggests that states, or planetary entities in this context, would act to prevent any single actor from attaining overwhelming dominance. Numerous examples of this can be seen through the history of alliances and conflict on Earth, and such patterns are likely to manifest in interplanetary relations. \citet{mearsheimer2001tragedy} has argued that great powers inherently seek to maximize their influence and security, often at the expense of others. This view applied to Mars would imply that Earth-based powers, which have significant investments in Martian settlements, may perceive a self-sufficient Mars as threatening their strategic interests.}

\rev{In the early phases of settlement, national governments and private entities on Earth will collaborate or compete in the creation of settlements on Mars, motivated by common scientific and economic interests. But as the martian settlements increase in population and begin to develop in situ resource extraction and industrial capabilities, their reliance on Earth will lessen. This would, in turn, lead to a power vacuum in which the competitive Earth powers will try either to establish their domination over Mars or prevent rival states from acquiring disproportionate influence. For example, privileged trade agreements or commercial bases on Mars would be subject to counterbalancing action by other Earth states, similar to those seen in traditional great power rivalries. Competition could intensify if Mars gains any degree of independence from Earth and starts asserting itself through political autonomy. As Mars grows as a power, logic \cite{mearsheimer2001tragedy} may see balancing behavior from the powers on Earth; for example, this could take the form of coalitional efforts to contain the power of Mars, or economic sanctions that prevent further technological and military development on Mars. As a result of this response from Earth, Mars might work harder toward forging interplanetary alliances.}

\rev{Hegemonic stability theory relies on the assumption that there can be one major power---the hegemon---capable of creating and maintaining order within the international system through the provision of public goods and enforcement of rules. The historical Pax Britannica and Pax Americana stand in testament to how hegemonic powers have been capable of facilitating economic and political stability. In a martian context, Earth as a whole---or one particular leading spacefaring nation, such as the United States or China---could operate as a hegemon during the early process of settlement. A hegemonic Earth would provide the critical infrastructure, technological expertise, and governance frameworks that would be necessary for the survival and stability of martian settlements. This would most likely be performed through the imposition of Earth-based legal and political norms on Mars, with settlements acting as an extension of terrestrial sovereignty. Such a system could foster cooperation by reducing the likelihood of inter-settlement conflicts on Mars and ensuring access to essential resources and supply chains. However, as Mars grows more self-reliant, the hegemonic order could face challenges. Settlers may perceive Earth’s dominance as exploitative, particularly if martian resources are extracted primarily for Earth’s benefit. This dynamic mirrors colonial relationships on Earth, where resource-rich peripheries often resisted the control of core states. The emphasis \citet{mearsheimer2001tragedy} places on the anarchic nature of international systems suggests that Mars, once capable, may seek to challenge Earth’s hegemony to assert its sovereignty. The sustainability of Earth's hegemonic role will depend on the balance between coercion and consent. Investment in the development of Mars and the political autonomy of settlers could buy cooperative relations and dampen the secessionist tendencies of martian settlements. However, a failure to adapt might lead to fragmentation, as martian settlements reject Earth's authority and pursue alternative governance structures.}

\rev{This discussion suggests that the binary division of "independence" and "subordination" in traditional sovereignty theory is limited in the context of interstellar governance. Mars settlement is particularly unique because human survival will depend on technological support from Earth, but the geographical isolation of living on Mars will give rise to a unique sense of community. This requires us to re-imagine the form of sovereignty---neither as a completely separated Earth jurisdiction nor a simple colonial dependency, but rather a dynamic balance of a "semi-sovereign" state. The devolution practice in Scotland offers important insights for the idea of semi-sovereignty: the Scotland Act of 1998, while maintaining the ultimate authority of the UK Parliament, transferred legislative power in areas such as education and healthcare to local Scottish authorities. This "partial delegation" model could be applied to space governance: for example, Earth-based nations could retain core responsibilities (such as compliance with interstellar treaties and maintenance of defense systems) but would gradually transfer authority over resource management and internal governance to the martian settlements. Any such effort would require a clear list of rights and responsibilities, and the establishment of a "permission unlocking" mechanism based on the maturity of the settlement---for instance, automatically granting the right to adjust local tax systems once a settlement has achieved stable operation of its oxygen circulation system for ten years.}

\rev{In the foreseeable future, Earth and Mars will most likely maintain a "symbiotic competition" relationship. The core value of the semi-independent model lies in providing an institutionalized container for this relationship: it acknowledges the primacy of Earth civilization while respecting the growth needs of the martian community. This exploration of dynamic balance may eventually give birth to humanity's first interplanetary federation---a political community that would allow different planetary civilizations to maintain their distinct characteristics while sharing sovereignty in specific areas. The success or failure of such an interplanetary political body is not only a problem limited to space governance but would also be an ultimate test of whether humanity can transcend geographical limitations and construct a model for the coexistence of diverse civilizations.}

Longer-term visions of space settlement could conceivably realize possibilities in which large settlements on Mars have achieved a degree of self-sustainability to the extent that political and economic independence from Earth is viable. In such cases, a settlement that declares itself independent would be claiming sovereignty over physical infrastructure and personnel that would otherwise be under the jurisdiction of the host state. The loss of such a long-term investment would likely be opposed by most host states, especially if no prior agreement had been determined such an act of succession. Drawing from the Catalan case study, if local separatist forces infringe upon national territorial sovereignty, then a state may leverage its comprehensive constitution—akin to Spain's Constitution or China's Constitution of the People's Republic of China—to invalidate the legitimacy of any succession effort, under the justification of ensuring territorial integrity. In such cases, conducting a referendum for independence loses its significance, only exacerbating regional unrest or leading to armed conflict after significant societal strife. For a martian independence referendum, the loss of such a political struggle would further complicate efforts at finding a balanced approach to shared sovereignty, as in the Quebec case study. Thus, to ensure that a referendum is well received, without a turbulent reaction or any violent effects, it is crucial to establish the conditions for its legality from the beginning.

\section{Conclusion}

This article examined the potential of using referendums on Mars for political decision making by analyzing three independence movements on Earth and analyzing possible analogies to situations that may arise on Mars. Actual attempts at Mars settlement may be decades or farther into the future, and many popular culture speculations about political futures in space tend to degrade into “space operas” with nation states as the primary actors. But pragmatic approaches toward shared sovereignty on Mars may have little in common with such speculative science fiction and instead can draw on lessons from history to explore plausible solutions that could function in the space environment. Even for long-term futures involving hypothetical cases of martian independence, any attempted solutions must satisfy the three “hard constraints” of technical capability, political feasibility, and long-term sustainability in order to be viable \cite{haqq2022sovereign,haqq2024constraints}. Such constraints likewise apply on Earth: the three case studies examined in this article demonstrate the extent to which a referendum could be politically feasible given the legal and militaristic realities that would be encountered, while complete economic self-sufficiency would also be required for secession efforts to be effective. On Mars, the technical capacity for life support further complicates such scenarios, which can introduce new problems in shared governance. Addressing such novel political problems on Mars is likely to introduce novel possibilities that have never arisen before.

The idea of a referendum on Mars to solve problems of shared sovereignty or self-determination also can be instructive for thinking about applying such tools on Earth. In most cases today, independence referendums are limited solely to the desires of ethnic groups for political self-determination; however, in the modern era of nation states that are often heterogeneously composed, this limitation on the use  of independence referendums lacks theoretical foundation and contemporary relevance \cite[see, e.g.,][]{shinoda2000re,zuo2014independence}. Adhering rigidly to an ethnic standard is, in essence, a veiled denial of the right of self-determination, attempting to stifle its legitimate exercise through a nebulous conceptual criterion. Mars, conversely, presents an opportunity to transcend this dilemma by broadening the scope of referendum eligibility. If planned efforts at space settlement were to explicitly extend the civic right of referendum to settlers on Mars, where no historical ethnic claims to sovereignty exist, then this could provide a mechanism for ensuring that Martian issues are resolved on Mars, without the intervention of Earth's international community or other external forces. At the same time, the availability of a referendum as a political tool, without any limitations based on the ethnicity of the settlers, would only serve to avert any further tendencies toward ethnic separatism on Mars. This, in turn, could reduce the likelihood of multiple competing political entities on Mars, leveraging referendums as a tool for negotiation and compromise between martian settlements and their host nations.

If Mars settlement is ultimately successful, then the generations of settlers will continue to develop a unique martian culture that may approach a size and complexity that necessitates more established local political structures. The use of referendums may be supplemented by other forms of governance that are suited to the needs of the martian settlements, which will likely draw upon ideas from Earth but ultimately will be adapted into solutions that work in the martian environment. We cannot easily extrapolate the likely political systems to develop on Mars from existing historical precedent, as Mars settlements will face numerous novel circumstances. Historically,  colonizing countries have exercised political leadership over their colonies through economic control or military force. But exerting such control will be even more difficult across interplanetary distances, with the biennial launch window and the exorbitant costs of space travel requiring martian settlers to develop at least some capabilities for self-reliance and self-governance. As such settlements on Mars continue to grow, and their connection with Earth dwindles, then new questions may arise regarding the extent to which referendums can be extended to encompass all Martian citizens. If Mars settlement is successful to the extent that populations of millions or more people are living in martian cities, then referendums themselves may be more limited in utility as other options become more pragmatic. It could be tempting from a modern terrestrial perspective to assume that an option like representative democracy or even a “stateless society” would be the most likely or desirable, but any governance structures that evolve on Mars will be uniquely adapted to the political needs of martian settlers, all of whom will depend upon life support technology in massive built environments to protect themselves from the harsh environment of space.

\section*{Acknowledgments}
This study was conducted during the 2024 Young Scientist Program (YSP) at the Blue Marble Space Institute of Science. This research did not receive any specific grant from funding agencies in the public, commercial, or not-for-profit sectors. Any opinions, findings, and conclusions or recommendations expressed in this material are those of the authors and do not necessarily reflect the views of any employer.

\bibliographystyle{elsarticle-num-names} 
\bibliography{main}

\end{document}